\documentclass[prper,aps,twocolumn,showpacs,groupedaddress,floats,final,superscriptaddress]{revtex4-1}
\usepackage{graphicx}
\usepackage{dcolumn}
\usepackage{bm}
\usepackage{color}
\usepackage{amsmath}
\usepackage{amssymb}
\usepackage{amsfonts}
\usepackage{enumitem}
\RequirePackage[
  hyperindex,colorlinks,bookmarksnumbered,
  plainpages=true,pdfstartview=FitH]{hyperref}
\hypersetup{linkcolor=blue,urlcolor=blue,citecolor=blue}
\usepackage{appendix}
\usepackage{hyperref}
\usepackage{soul} % \st{Hello world} to strickethrough

\newcommand{\h}{\mathit{\boldsymbol{h}}}
\newcommand{\X}{\mathit{\boldsymbol{x}}}
\newcommand{\A}{{\bf a}}
\newcommand{\B}{{\bf b}}

\newcommand{\bkappa}{\boldsymbol{\kappa}}

\newcommand{\loc}{\mathrm{loc}}
\newcommand{\opt}{\mathrm{opt}}

\newcommand{\cl}{\mathrm{cl}}

\newcommand{\xv}{\mathit{\boldsymbol{y}}}
\newcommand{\beq}{\begin{equation}}
\newcommand{\eeq}{\end{equation}}
\newcommand{\ba}{\begin{array}}
\newcommand{\ea}{\end{array}}
\newcommand{\bea}{\begin{eqnarray}}
\newcommand{\eea}{\end{eqnarray}}

\begin{document}

\title{Projective quantum Monte Carlo simulations guided by unrestricted neural network states}

%%%%%%%%%%%%%%%%%%%%%%%%%%%%%%%%%%%%%%%%%%%%%%%%%%%%%%%%%%%%%%
\author{E. M. Inack}
\affiliation{The Abdus Salam International Centre for Theoretical Physics, 34151 Trieste, Italy}
\affiliation{SISSA - International School for Advanced Studies, 34136 Trieste, Italy}
\affiliation{INFN, Sezione di Trieste, 34136 Trieste, Italy} 
\author{G. E. Santoro}
\affiliation{The Abdus Salam International Centre for Theoretical Physics, 34151 Trieste, Italy}
\affiliation{SISSA - International School for Advanced Studies, 34136 Trieste, Italy}
\affiliation{CNR-IOM  Democritos  National  Simulation  Center,  Via  Bonomea  265,  34136  Trieste,
Italy} 
\author{L. Dell'Anna}
\affiliation{Dipartimento di Fisica e Astronomia ``Galileo Galilei", Universit{\`a} di Padova, 35131 Padova, Italy}
\affiliation{CNISM, Sezione di Padova, Italy}
\author{S. Pilati}
\affiliation{School of Science and Technology, Physics Division, Universit{\`a}  di Camerino, 62032 Camerino (MC), Italy}
%%%%%%%%%%%%%%%%%%%%%%%%%%%%%%%%%%%%%%%%%%%%%%%%%%%%%%%%%%%

\begin{abstract}
We investigate the use of variational wave-functions that mimic stochastic recurrent neural networks, specifically, unrestricted Boltzmann machines, 
as guiding functions in projective quantum Monte Carlo (PQMC) simulations of quantum spin models.
As a preliminary step, we investigate the accuracy of such  unrestricted neural network states as 
variational Ans\"atze for the ground state of the ferromagnetic quantum Ising chain. 
We find that by optimizing just three variational parameters, independently on the system size, accurate ground-state energies are obtained, comparable to those previously obtained using restricted Boltzmann machines with few variational parameters per spin. 
Chiefly, we show that if one uses optimized unrestricted neural network states as guiding functions for importance sampling the efficiency of the PQMC algorithms is greatly enhanced, 
drastically reducing the most relevant systematic bias, namely that due to the finite random-walker population. The scaling of the computational cost with the system size changes from the exponential scaling characteristic of PQMC simulations performed without importance sampling, to a polynomial scaling, even at the ferromagnetic quantum critical point.
The important role of the protocol chosen to sample hidden-spins configurations, in particular at the critical point, is analyzed.
 We discuss the implications of these findings for what concerns the problem of simulating adiabatic quantum  optimization  using stochastic algorithms on classical computers.
\end{abstract}

\maketitle
\section{Introduction}
Quantum Monte Carlo (QMC) algorithms are generally believed to be capable of predicting equilibrium 
properties of quantum many-body systems at an affordable computational cost, even for relatively large system sizes, at 
least when the sign problem does not occur.
However, it has recently been shown that the computational cost to simulate the ground state of a quantum Ising model with a 
simple projective QMC (PQMC) algorithm that does not exploit importance sampling techniques scales exponentially with the system size,
 making large-scale simulations unfeasible~\cite{inack2}. This happens in spite of the fact that the Hamiltonian is sign-problem free.
PQMC methods have found vast use in condensed matter physics, in chemistry, and beyond (see, e.g., Refs.~\cite{ceperley1986quantum, hammond1994monte, foulkes2001quantum, carlson2015quantum}). 
Shedding light on their computational complexity, and possibly improving it by using importance sampling techniques based on 
novel variational wave-functions, are therefore very important tasks. We address them in this Article.

PQMC algorithms have 
recently emerged as useful computational tools also to investigate the potential efficiency of adiabatic quantum computers in 
 solving large-scale optimization problems via quantum annealing~\cite{Finnila_CPL94,santorotheory,boixo2014evidence,inack,troyerheim}. 
In particular, it has been shown that the (stochastic) dynamics of simple PQMC simulations allows to tunnel through tall barriers of (effectively) double-well  models even more efficiently than an adiabatic quantum computer which exploits incoherent quantum tunneling~\cite{isakovtunneling,jianginstanton,mazzolaquantumchemistry,inack2}. 
This result seems to suggest that there might be no systematic quantum speed-up in using a quantum annealing device 
to solve an optimization problem, compared to a stochastic QMC simulation performed on a classical computer~\cite{isakovtunneling}.
Remarkably,  this computational advantage of the PQMC simulations with respect to the expected behavior of a quantum annealing device occurs also in more challenging models with frustrated couplings~\cite{inack2}, as in the recently introduced Shamrock model, where QMC algorithms based on the (finite temperature) path-integral formalism display instead an exponential slowdown of the tunneling dynamics~\cite{aminshamrock}.
 This result further stresses the importance of 
 shedding light on the computational complexity of PQMC algorithms: if these computational techniques allowed one to simulate, with a polynomially scaling computational cost, both the ground-state properties of 
 a model Hamiltonian, and also the tunneling dynamics of a quantum annealing device described by such Hamiltonian~\cite{inack2}, then the quantum speedup mentioned above 
 would be very unlikely to be achieved. We focus in this paper on the first of the two aspects, specifically, 
 on analyzing and improving the scaling of the computational cost to simulate  ground-state properties of quantum Ising models.

It is well known that the efficiency of PQMC algorithms can be enhanced by implementing importance sampling techniques
 using as guiding functions accurate variational Ans\"atze~\cite{foulkes2001quantum}. However, building accurate variational wave-functions for generic many-body systems is 
 a highly non trivial task. Recently, variational wave-functions that mimic the structure of neural networks have 
 been shown to accurately describe ground-state properties of quantum spin and lattice models~\cite{carleotroyer,saito,saito2}. 
The representational power and the entanglement content of such variational states, now referred to as neural network states, 
have been investigated~\cite{deng2017quantum,chen2018equivalence,glasser2018neural,gao2017efficient,freitas2018neural}, 
showing, among other properties, that they are capable of describing volume-law entanglement.
The authors of Ref.~\cite{carleotroyer} considered neural network states that mimic restricted Boltzmann machines (RBM), i.e. such that no interaction among hidden spins is allowed.
One very appealing feature of such restricted neural network states is that the role of the hidden spins can be accounted for analytically, 
without the need of Monte Carlo sampling over hidden variables.
Furthermore, such states provide very accurate ground-state energy predictions, which can be systematically improved by increasing the number of hidden spins per visible spin (later on referred to as hidden-spin density). However, this high accuracy is obtained at the cost of 
optimizing a number of variational parameters that increases with the system size. 
This optimization task can be tackled using powerful optimization algorithms such as the stochastic reconfiguration method (see, e.g, Ref.~\cite{stocric}). Yet, 
having to optimize a large number of variational parameters is not desirable in the context of quantum annealing simulations, since one would be dealing with a variational optimization problem, potentially even more difficult than the original classical optimization problem. 

In this Article, we consider instead neural network states that mimic unrestricted Boltzmann machines (uRBMs), allowing intra-layer correlations 
among hidden spins, beyond the inter-layer hidden-visible correlations and the intra-layer visible-visible correlations (see Fig.~\ref{fig0}). The structure of these states resembles the one of the shadow wave functions originally introduced to describe 
quantum fluids and solids~\cite{shadowvitiello,reatto1988shadow}. We test their representational power considering as a testbed the ferromagnetic quantum Ising chain. 
We find that by optimizing just three  variational parameters, independently on the system size, very accurate ground-state energies are obtained, comparable to the case of restricted neural network states with one hidden spin per visible spin.
Such a small number of variational parameters is a particularly appealing feature in the context of quantum annealing problems. However, it comes 
at the prize of having to perform Monte Carlo sampling over hidden-spin configurations.

The main goal of this Article is to show that the above-mentioned unrestricted neural network states can be used as a guide for 
importance sampling in PQMC simulations. This also implies that the development of  
neural network states 
can be limited to obtaining reasonably accurate, but not necessarily exact, variational Ans\"atze, since the residual error can be eliminated 
within the PQMC simulation. 
In particular, we provide numerical evidence that the major source of systematic bias of 
the PQMC algorithms, 
namely the bias originating from the finite size of the random-walker population which has to be stochastically 
evolved in any PQMC simulation, can be 
drastically reduced using optimized unrestricted neural network states, even at the point of changing the scaling of the required population size 
from exponential (corresponding to the case without importance sampling) to polynomial in the system size. 
This also implies a change of computational complexity from exponential to polynomial. For comparison, we  show that a conventional variational wave-function of the Boltzmann type (with no hidden spins), instead, does not determine a comparable efficiency improvement. 

The rest of the Article is organized as follows: in Section~\ref{secvar} we define the conventional Boltzmann-type variational wave functions and 
the unrestricted neural network states, and we then analyze how accurately they predict the ground-state energy of the quantum Ising chain via optimization of, respectively one and three, variational parameters. Section~\ref{secgfmc} deals with the continuous-time PQMC algorithm and with the implementation 
of importance sampling using both Boltzmann-type wave functions and, chiefly, unrestricted neural network states, showing how the systematic bias 
due to the finite random-walker population is affected, both at and away from the quantum critical point. The important effect of choosing different sampling protocols for the hidden spins is also analyzed.
Our conclusions and the outlook are reported in Section~\ref{secconc}.

\section{Unrestricted neural network states for quantum Ising models}
\label{secvar}
In this article, we consider as a test bed the one-dimensional ferromagnetic quantum Ising Hamiltonian:
\begin{equation}
\hat{H}=\hat{H}_{\cl}+\hat{H}_{\mathrm{kin}},
\label{H}
\end{equation}
where $\hat{H}_{\cl}=-J\sum_{i=1}^N {\sigma}^{z}_{i} {\sigma}^{z}_{i+1}$ and $\hat{H}_{\mathrm{kin}}=-\Gamma \sum_{i=1}^{N} {\sigma}^{x}_{i}$. $\sigma^x_i$, $\sigma^y_i$, and $\sigma^z_i$ indicate Pauli matrices acting on spins at the lattice site $i$. $N$ is the total number of spins, and we adopt periodic boundary conditions, i.e. $ {\sigma}^{\alpha}_{N+1}={\sigma}^{\alpha}_{1}$, with $\alpha=x,y,z$. The parameter $J>0$ fixes the strength of the ferromagnetic interactions among nearest-neighbor spins.  In the following, we set $J=1$. All energy scales are henceforth expressed in units of $J$. The parameter  $\Gamma$ fixes the intensity of a transverse magnetic field. Given $\left| x_i \right>$ an eigenstate of the Pauli matrix ${\sigma^z_i}$ with eigenvalue $x_i=1$ when $\left|x\right>=\left|\uparrow \right>$ and $x_i=-1$ when $\left|x \right>=\left|\downarrow \right>$, the quantum state of $N$ spins is indicated by $\left|\X \right> = \left|x_1 x_2 ... x_N\right>$. 
Notice that the function $E_{\cl}(\X)=\langle \X|  \hat{H}_{\cl}| \X \rangle$ (with $\X=\left(x_1,x_2,\dots,x_N\right)$) corresponds to the Hamiltonian function of a classical Ising model, while the operator $\hat{H}_{\mathrm{kin}}$ introduces quantum (kinetic) fluctuations.

Our first goal is to develop trial wave functions that closely approximate the ground state wave function $\Psi_{0}{(\X)}=\left< \X | \Psi_0\right>$ of the Hamiltonian~(\ref{H}).  
A simple Ansatz can be defined as 
\beq
\Psi_{\bkappa}{(\X)}=e^{-\beta E_{\cl} ( {\X} )  } = e^{-K_{1}\sum_{i=1}^{N}x_i x_{i+1} }  \;.
\eeq 
$\bkappa$ is here a set of real variational parameters to be optimized. 
Their values are obtained by minimizing the average of the energy, as in standard variational quantum Monte Carlo approaches. 
In this case, only one parameter $K_1=\beta $ is present, $\bkappa=\{ K_1 \}$. 
This  choice is inspired by the classical Boltzmann distribution where $\beta$ would play the role of a fictitious inverse temperature. 
The above Ansatz will be referred to as Boltzmann-type wave function. 

\begin{figure}
\begin{center}
\includegraphics[width=1.0\columnwidth]{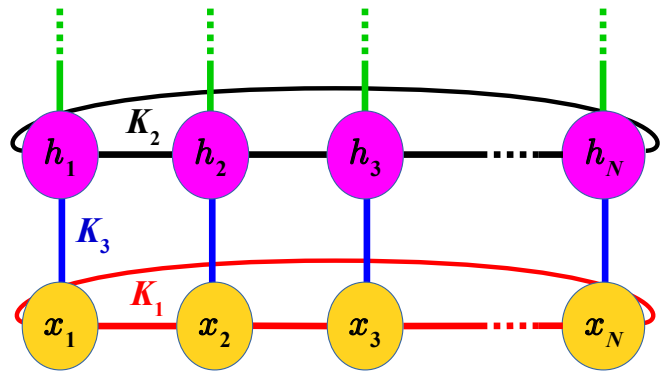}
\caption{(color online). Structure of the unrestricted Boltzmann machine. The lower (yellow) nodes depict visible spins, the upper (magenta) nodes depict the hidden spins. The horizontal segments indicate intralayer visible-visible and hidden-hidden correlations. The vertical (blue) segments represent the interlayer correlations between the corresponding visible and hidden spins. The green lines allude to a possible extension to deep layers architectures.
}
\label{fig0}
\end{center}
\end{figure}

A more sophisticated Ansatz can be constructed by using a generative stochastic artificial neural network, namely an uRBM (see Fig.~\ref{fig0}). 
Beyond the visible spin variables $\X=\left( x_1,x_2,\dots,x_N\right)$, one introduces $N$ hidden spin variables $\h=\left( h_1,h_2,\dots,h_N\right)$, 
taking values $h_i= \pm 1$ (with $i=1,\dots,N$). 
Periodic boundary conditions within the layers are also incorporated, i.e $x_{N+1}=x_1$ and $h_{N+1}=h_1$. 
The trial wave function is thus written in the following integral form:
\beq \label{ann}
\Psi_{\bkappa}(\X)=\sum_{\h} \phi_{\bkappa}\left(\X,\h \right) \;,
\eeq
where,
\beq
 \phi_{\bkappa}(\X,\h)=e^{-\sum_{i=1}^{N} \left( K_{1} x_i x_{i+1} +K_{2} h_i h_{i+1}+K_{3}x_i h_i \right) } \;. 
\label{ann1}
\eeq
Notice that the architecture of this uRBM  includes correlations  between nearest-neighbor visible spins, between nearest-neighbor hidden spins, as well as between pairs of  visible and hidden spins with the same index $i$. These three correlations are parametrized by the three constants $K_1$, $K_2$, and $K_3$, respectively. 
With this uRBM trial Ansatz, the set of variational parameters is $\bkappa=\{K_1,K_2,K_3  \}$. 
It is straightforward to generalize the uRBM Ansatz including more layers of hidden spins. Every additional hidden-spin layer adds two more variational parameters, and it effectively represents the application of an imaginary-time Suzuki-Trotter step $e^{-\Delta \tau \hat{H}}$ for a certain time step $\Delta \tau$. Thus, a deep neural network state with many hidden layers can represent a long imaginary-time dynamics, which projects out the ground state provided that the initial state is not orthogonal to it. In fact, the mapping between deep neural networks and the imaginary time projection has been exploited in Refs.~\cite{carleo2018constructing,freitas2018neural} to construct more complex neural network states.
In this article we consider only the single hidden-spin layer uRBM, since this Ansatz turns out to be adequate for the ferromagnetic quantum Ising chain. 
The multi hidden-spin layer Ansatz might be useful to address more complex models as, e.g, frustrated Ising spin glasses. Extensions along these lines are left as future work. 

In a recent work~\cite{carleotroyer}, Carleo and Troyer considered a restricted Boltzmann machine (RBM), where direct correlations among hidden spins were not allowed. Their Ansatz included a larger number of hidden spins, as well as more connections between visible and hidden spins, leading to an extensive number of variational parameter proportional to $\alpha N$, where $\alpha=1,2,\dots$. 
One advantage of the RBM, due to the absence of hidden-hidden correlations, is that the role of hidden spins can be analytically traced out. The uRBM we employ, which is analogous to the shadow wave functions used to describe quantum fluid and solids, includes only three variational parameters, independently of the system size. However, their effect has to be addressed by performing sampling of hidden spins configurations, as described below.
 It is worth pointing out that correlations beyond nearest-neighbor spins could also be included in the uRBM Ansatz, with straightforward modifications in the sampling algorithms described below. We mention here also that, as shown in Ref.~\cite{gao2017efficient}, neural network states with intra-layer correlations can be mapped to deep neural networks with more hidden layers, but no intra-layer correlations.

%
%%%%%%%%%%%%%%%%%%%%%%%%%%%%%%%%%%%%%%%%%%%%%%%%%%%%%%%%%%%%%%%%%%%%%
%%%   figure 0
%%%%%%%%%%%%%%%%
%%%%%%%%%%%%%%%%%%%%%%%%%%%%%%%%%%%%%%%%%%%%%%%%%%%%%%%%%%%%%%%%%%%%%
%%%%%%%%%%%%%%
%

In the case of an uRBM variational wave function, the average value of the energy $E={\langle \hat{H} \rangle}$ is computed as follows
\bea
 {\langle \hat{H} \rangle} &=&  \frac{\langle \Psi_{\bkappa} |\hat{H}|\Psi_{\bkappa} \rangle}{\langle \Psi_{\bkappa} |\Psi_{\bkappa} \rangle} 
 = \frac{\sum_{\X,\X'}  \Psi_{\bkappa}(\X) H_{\X,\X'}  \Psi_{\bkappa}(\X')  }{\sum_{\X}  \Psi_{\bkappa}(\X)  \Psi_{\bkappa}(\X) }  \nonumber \\
 &=& \frac{\sum_{\X,\X'}  \left[ \sum_{\h_\A} \phi_{\bkappa}(\X,\h_\A) \right] H_{\X,\X'}  \left[ \sum_{\h_\B} \phi_{\bkappa}(\X',\h_\B) \right]  }{\sum_{\X}  \left[ \sum_{\h_\A} \phi_{\bkappa}(\X,\h_\A) \right] \left[ \sum_{\h_\B} \phi_{\bkappa}(\X,\h_\B) \right] }  \nonumber \\
&= & \ll E_{\loc}( \X, \h_\B )\gg \;,
\label{eng}
\eea
where the local energy $E_{\loc}(\X,\h)$ is defined as 
\beq
\label{elocal}
E_{\loc}(\X,\h)=\frac{\sum_{\X'} H_{\X,\X'}   \phi_{\bkappa}(\X',\h)} {   \phi_{\bkappa}(\X,\h)} \;,
\eeq
with $H_{\X,\X'}=\langle \X|\hat{H}|\X^\prime\rangle$. $\h_\A$ and $\h_\B $ indicate two hidden spin configurations. Notice that the formula for the local energy can be symmetrized with respect to the two sets of hidden spins $\h_\A$ and $\h_\B$, providing results with slightly reduced statistical fluctuations. The double brackets $\ll \cdots \gg$ indicate the expectation value over the visible-spin configurations $\X$ and two sets of hidden spins configurations 
$\h_\A$ and $\h_\B$, sampled from the following normalized probability distribution:
\beq
p(\X,\h_\A,\h_\B)=\frac{\phi_{\bkappa}(\X,\h_\A)  \phi_{\bkappa}(\X,\h_\B) }{\sum_{\X,\h_\A,\h_\B}   \phi_{\bkappa}(\X,\h_\A) \phi_{\bkappa}(\X,\h_\B)} \;.
\eeq
As in standard  Monte Carlo approaches, this expectation value is estimated as the average of $E_{\loc}(\X,\h)$ over a (large) set of uncorrelated configurations, sampled according to $p(\X,\h_\A,\h_\B)$. The statistical uncertainty can be reduced at will by increasing the number of sampled configurations.
The optimal variational parameters $\bkappa_{\opt}$ that minimize the energy expectation value can be found using a stochastic optimization method. 
We adopt a relatively simple yet quite efficient one, namely the stochastic gradient descent algorithm (see, e.g.,~\cite{becca_sorella}). While more sophisticated algorithms exist as, e.g., the stochastic reconfiguration method~\cite{stocric}, such methods are not necessary here since the Ans\"atze that we consider include a very small number of variational parameters, one or three. In fact, in these cases the optimal variational parameters can be obtained also by performing a scan on a fine grid. By doing so, we obtain essentially the same results provided by the stochastic gradient descent algorithm.

We assess the accuracy of the optimized variational wave functions by calculating the relative error 
\begin{equation} \label{eqn:erel}
e_{\mathrm{rel}} = \frac{\left|E - E_{\mathrm{JW}}\right|}{\left|E_{\mathrm{JW}}\right|} \;,
\end{equation} 
in the obtained variational estimate $E$ of the ground state energy of the Hamiltonian in Eq.~(\ref{H}). 
$E_{\mathrm{JW}}$ is the exact finite size ground state energy of the quantum Ising chain. 
It is obtained by performing the Jordan--Wigner transformation, followed by a Fourier and the Bogoliubov transformations. 

\begin{figure}
\begin{center}
\includegraphics[width=1.0\columnwidth]{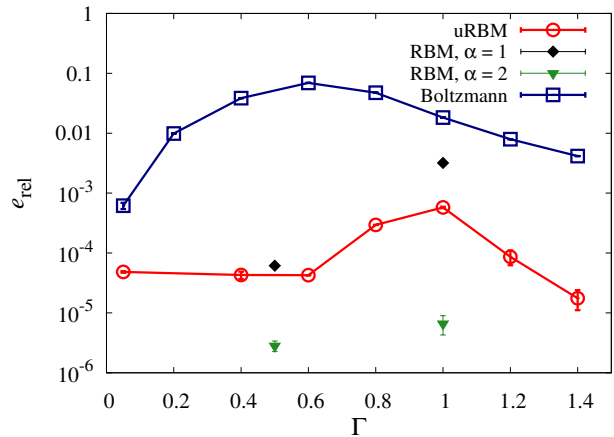}
\caption{(color online). Relative error $e_{\mathrm{rel}}$ in the variational estimates of the ground-state energy, see Eq.~\eqref{eqn:erel}, 
as a function of the transverse field $\Gamma$, obtained using the simple Boltzmann wave function and for the unrestricted Boltzmann machine (uRBM) Ansatz. 
The system size is $N=80$. For comparison, we also show the data corresponding to the restricted Boltzmann machine (RBM) from Ref.~\onlinecite{carleotroyer}, 
where $\alpha$ indicates the hidden-spin density.  The thin lines are guides to the eyes.
}
\label{fig1}
\end{center}
\end{figure}

Figure~\ref{fig1} displays the relative error $e_{\mathrm{rel}}$ in Eq.~\eqref{eqn:erel} corresponding to the variational wave functions introduced above, 
as a function of the transverse field $\Gamma$. The system size is $N=80$, which is here representative of the thermodynamic limit.
The Boltzmann-type Ansatz does not provide particularly accurate predictions. 
In the ferromagnetic phase $\Gamma<1$, the relative error is up to $10\%$.
The uRBM, instead, provides very accurate predictions. The relative error is always below $0.1\%$. 
The largest discrepancy occurs at the quantum critical point $\Gamma=1$. 
Such high accuracy is remarkable, considering that the uRBM Ansatz involves only $3$ variational parameters. 
It is also worth mentioning that very similar accuracies are obtained also for different system sizes.
Therefore, the uRBM Ansatz represents a promising guiding function for simulations of quantum annealing optimization of disordered models. 
As a term of comparison, we show in Fig.~\ref{fig1} the results obtained in Ref.~\onlinecite{carleotroyer} using the RBM Ansatz. 
The relative errors corresponding to the RBM with hidden-unit density $\alpha=1$ are larger than those corresponding to the uRBM, despite the fact that the 
RBM Ansatz involves a larger number of variational parameters. 
However, it is worth stressing that the RBM results can be systematically improved by increasing $\alpha$. 
For example, with $\alpha=2$ the RBM relative errors  are approximately an order of magnitude smaller than those corresponding to the uRBM Ansatz. 

%%%%%%%%%%%%%%%%%%%%%%%%%%%%%%%%%%%%%%%%%%%%%%%%%%%%%%%%%%%%%%%%%%%%%
%%%   figure 1
%%%%%%%%%%%%%%%%
%%%%%%%%%%%%%%%%%%%%%%%%%%%%%%%%%%%%%%%%%%%%%%%%%%%%%%%%%%%%%%%%%%%
%%%%%%%%%%%%%%%%%%%%%%%%%%%%%%%%%%%%%%%%%%%%%%%%%%%%%%%%%%%%%%%%%%%

\section{Importance sampling guided by unrestricted neural network states} \label{secgfmc}
In this section we discuss how optimized variational wave functions can be utilized to boost the performance of PQMC simulations. 
First, we consider the implementation of the PQMC algorithm without guiding functions. PQMC methods allow one to extract ground-state properties of quantum many-body systems~\cite{andersonJB,kalos} by stochastically simulating the Schr\"odinger equation in imaginary-time $\tau=i t$. 
In the Dirac notation, this equation is written as:
\begin{eqnarray}
 -\frac{\partial}{\partial \tau}|\Psi(\tau)\rangle=(\hat{H}-E_{\mathrm{ref}})|\Psi(\tau)\rangle.
\label{DiracSHE}
\end{eqnarray} 
The reduced Planck constant is set to $\hbar=1$ throughout this Article. $E_{\mathrm{ref}}$ is a reference energy introduced to stabilize the simulation, as discussed later.
Eq.~(\ref{DiracSHE}) is simulated by iteratively applying the equation 
$\Psi(\X,\tau+\Delta \tau)=\sum_{\X^\prime}G(\X,\X^\prime,\Delta \tau)\Psi(\X^\prime,\tau)$. 
$\Delta\tau$ is a (short) time step and $G({\X},{\X}^\prime,\Delta \tau)= \langle \X|e^{-\Delta \tau(\hat{H}-E_{\mathrm{ref}})} |\X^\prime\rangle$ is the Green's function of Eq.~(\ref{DiracSHE}). Below it is discussed how one can write a suitable explicit expression.
Long propagation times $\tau$ are achieved by iterating many (small) time steps $\Delta \tau$, allowing one to sample, in the $\tau \rightarrow \infty$ limit, spin configurations with a probability density proportional to the ground state wave function $\Psi_0(\X)$ (assumed to be real and non negative).
One should notice that the Green's function $G({\X},{\X}^\prime,\Delta \tau)$ does not define a stochastic matrix; while its elements are nonnegative, one has $\sum_{\X}G({\X},{\X}^\prime,\Delta \tau)\neq 1$, in general. Therefore, it cannot be utilized to define the transition matrix of a conventional Markov chain Monte Carlo simulation. This problem can be circumvented by rewriting the Green's function as $G({\X},{\X}^\prime,\Delta \tau)=G_{\mathrm{T}}({\X},{\X}^\prime,\Delta \tau)b_{\X^\prime}$, where $G_{\mathrm{T}}({\X},{\X}^\prime,\Delta \tau)$ is by definition stochastic, and the normalization factor is $b_{\X^\prime} = \sum_{\X}G({\X},{\X}^\prime,\Delta \tau)$. 
A stochastic process can then be implemented, where a large population of equivalent copies of the system, in jargon called {\em walkers}, is evolved. Each walker represents one possible spin configuration ${\X}_n^\prime$ (the index $n$ labels different walkers), and is gradually modified by performing spin-configuration updates  according to $G_{\mathrm{T}}({\X}_n,{\X}_n^\prime,\Delta \tau)$. Thereafter, their (relative) weights $w_n$ are accumulated according to the rule $w_n\rightarrow w_n b_{\X_n^\prime}$, starting with equal initial weights $w_n=1$ for all the walkers in the initial population.
While this implementation is in principle correct, it is known to lead to an exponentially fast signal loss as the number of Monte Carlo steps increases. This is due to the fact that the relative weight of few walkers quickly becomes dominant, while most other walkers give a negligible contribution to the signal.
An effective remedy consists in introducing a branching process, where each walker is replicated (or annihilated) a number of times corresponding, on average, to the weight $w_n$. 
The simplest correct rule consists in generating, for each walker in the population at a certain imaginary time $\tau$, a number of descendants $n_{\mathrm{d}}$ in the population at imaginary time $\tau+\Delta \tau$. $n_{\mathrm{d}}$ is defined as $\mathrm{int}\left[w_n+\eta\right]$, where $\eta\in \left[0,1\right]$ is a uniform random number, and the function $\mathrm{int}\left[\right]$ gives the integer part of the argument~\cite{thijssen}. Clearly, after branching has been performed, all walkers have the same weight $w_n=1$.
Therefore, the number of walkers in the population fluctuates at each PQMC iteration and can be kept  close to a target value by adjusting the reference energy $E_{\mathrm{ref}}$.
Introducing the branching process provides one with a feasible, possibly efficient algorithm. However, such as process might actually introduce a systematic bias if the average population size $N_w$ is not large enough. The bias originates from the spurious correlations among walkers generated from the same ancestor~\cite{becca_sorella}. This effect becomes negligible in the $N_w\rightarrow \infty$ limit, but might be sizable for finite $N_w$. It is known to be the most relevant and subtle possible source of systematic errors in PQMC algorithms~\cite{nemec,boninsegnimoroni,pollet2018stochastic}. 
In fact, it was shown in Ref.~\cite{inack2} that in order to determine with a fixed target relative error, the ground state energy of the ferromagnetic quantum Ising chain with the (simple) diffusion Monte Carlo algorithm (which belongs to the category of PQMC methods), the walker-population size $N_w$ has to exponentially increase with the system size $N$. This implies an exponentially scaling computational cost.
\begin{figure}
\begin{center}
\includegraphics[width=1.0\columnwidth]{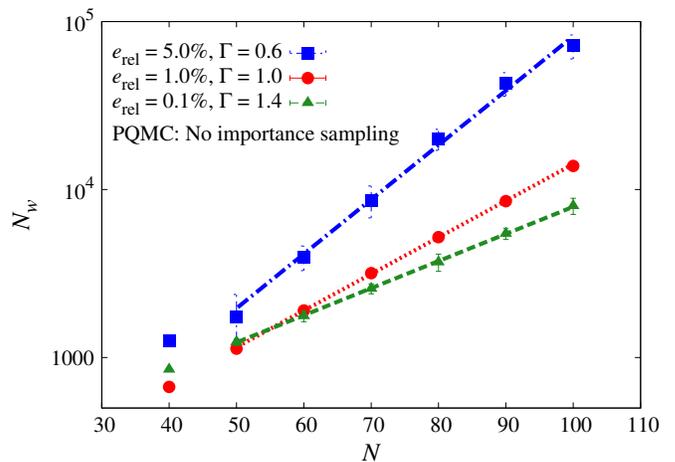}
\caption{(color online). Number of random walkers $N_w$ required to determine, using the PQMC algorithm without importance sampling, the ground-state energy 
with a relative error $e_{\mathrm{rel}}$, see Eq.~\eqref{eqn:erel}, as a function of the system size $N$. 
Different datasets correspond to different transverse field intensities $\Gamma$ and different relative errors. The lines represent exponential fitting functions.
}
\label{fig2}
\end{center}
\end{figure}
%%%%%%%%%%%%%%%%%%%%%%%%%%%%%%%%%%%%%%%%%%%%%%%%%%%%%%%%%%%%%%%%%%%%%

A promising strategy to circumvent the aforementioned problem is to introduce the so-called importance sampling technique. This is indeed a well established approach  to boost the efficiency of  PQMC simulations (see, e.g, Ref.~\cite{foulkes2001quantum}) because it has the potential to reduce the number of walkers needed to attain a given accuracy~\cite{becca_sorella}. It consists in evolving a function $f({\X},\tau)=\Psi ({\X},\tau) \psi_T({\X})$ via a modified imaginary-time Schr\"odinger equation. $\psi_T({\X})$ is a guiding function designed to accurately approximate the ground-state wave function. Its role is to  favor the sampling of  configurations with high probability amplitude. The obtained modified imaginary-time Schr\"odinger equation is solved via a Markov process defined by the following equation:
\begin{equation}
\label{masterf}
f(\X,\tau+\Delta \tau) = \sum_{\X^\prime} \tilde{G}(\X,\X^\prime,\Delta \tau) f(\X^\prime,\tau), 
\end{equation}
where the modified Green's function is given by $\tilde{G}({\X},{\X}^\prime,\Delta \tau)= G(\X,\X^\prime,\Delta \tau)\frac{\psi_T({\X})}{\psi_T({\X^\prime})}$.
A suitable approximation for the modified Green's function can be obtained by dividing the time step $\Delta \tau$ into $M$ shorter time steps 
$\delta \tau=\Delta \tau/M$. If $\delta \tau$ is sufficiently short, one can employ a Taylor expansion truncated at the linear term, 
$\tilde{G}({\X},{\X}^\prime,\Delta \tau) \cong \left[\tilde{g}({\X},{\X}^\prime,\delta \tau)\right]^M$, where:
\begin{equation}
\tilde{g}({\X},{\X}^\prime,\delta \tau) =\big[ \delta_{\X,\X^\prime}-  \delta \tau (H_{\X,\X^\prime}-E_{\mathrm{ref}}\delta_{\X,\X^\prime}) \big]\frac{\psi_T({\X})}{\psi_T({\X^\prime})} \;.
\end{equation} 
With this approximation, Eq.~\eqref{masterf} defines a stochastic implementation of the power method of linear algebra. 
Convergence to the exact ground state is guaranteed as long as $\delta \tau$ is smaller than a finite value, sufficiently small to ensure that all matrix elements of 
$\tilde{g}({\X},{\X}^\prime,\delta \tau)$ are not negative~\cite{schmidt2005green}. 
As the system size increases, shorter and shorter time steps $\delta \tau$ are required. This leads to pathologically inefficient simulations, since in this regime the identity operator dominates, resulting in extremely long autocorrelation times.
This problem can be solved by adopting the continuous-time Green's function Monte Carlo (CTGFMC) algorithm. The derivation and the details of this algorithm are given in Ref.~\cite{becca_sorella,SorellaCTGFMC}, and so we only sketch it here. The idea is to formally take the $M\rightarrow \infty$ limit, and determine the (stochastic)  time interval $\delta \tau^\prime$ that passes before the next configuration update occurs. It is convenient to bookkeep the remaining time $\delta \tau_t$ left to complete a total interval of time $\Delta \tau$. This is to ensure that each iteration of the PQMC simulation corresponds to a time step of duration  $\Delta \tau$. 
The time interval  $\delta \tau^\prime$ is sampled using the formula $\delta \tau^\prime=\mathrm{Min} \big( \delta \tau_t, \frac{\ln(1-\xi)}{E_{\loc}(\X')-E_{\cl}(\X')}  \big)$ with $\xi\in(0,1)$ being a uniform random number.
The spin-configuration update $\X^\prime \rightarrow \X$ (with $\X^\prime \neq \X$) is randomly selected from the probability distribution  
\begin{equation} \label{eqn:txx}
\begin{array}{ll}
t_{{\X},{\X}^\prime} &= \displaystyle \frac{p_{{\X},{\X}^\prime}}{\sum_{\X \neq \X^\prime}p_{{\X},{\X}^\prime }} \vspace{2mm} \\
p_{{\X},{\X}^\prime} &= \displaystyle \frac{\tilde{g}({\X},{\X}^\prime,\delta \tau')}{\sum_{\X} \tilde{g}({\X},{\X}^\prime,\delta \tau')}
\end{array} \;.
\end{equation}
Notice that, with the Hamiltonian (\ref{H}), $\X$ differs from $\X^\prime$ only for one spin flip.
%
%$\delta \tau^\prime$ being small, 
The weight-update factor for the branching process takes the exponential form $b_{\X'} = e^{-\delta \tau^\prime [E_{\loc}(\X') -E_{\mathrm{ref}} ]}$, where the local energy is now $E_{\loc}(\X')=\sum_{\X} H_{\X,\X^\prime}\frac{\psi_T({\X})}{\psi_T({\X^\prime})} $.

In summary, the CTGFMC algorithm requires to perform, for each walker $n$ in the population, the following steps:
\begin{description}
\item[i)] initialize the time interval $\delta \tau_t=\Delta \tau$, and the weight factor $w_n=1$; 
\item[ii)] sample the time $\delta\tau^\prime$ at which the the configuration update $\X^\prime \rightarrow \X$ might occur;
\item[iii)] if $\delta \tau^\prime <\delta \tau_t$, update $\X'$ with a transition probability $t_{{\X},{\X}^\prime}$ in Eq.~\eqref{eqn:txx}, 
else set $\delta \tau^\prime = \delta \tau_t$;
\item[iv)] accumulate the weight factor according to the rule $w_n \rightarrow w_n b_{\X'}$ and set $\delta \tau_t  \rightarrow \delta \tau_t -\delta \tau^\prime$;
\item[v)] Go back to step {\bf ii)} until $\delta \tau_t =0$; 
\item[vi)] finally, perform branching according to the total accumulated weight factor $w_n$.
\end{description}
This continuous-time algorithm implicitly implements the exact imaginary-time modified Green's function $\tilde{G}({\X},{\X}^\prime,\Delta \tau)$. 

%%%%%%%%%%%%%%%%%%%%%%%%%%%%%%%%%%%%%%%%%%%%%%%%%%%%%%%%%%%%%%%%%%%%%
%%%   figure 3
%%%%%%%%%%%%%%%%
\begin{figure}
\begin{center}
\includegraphics[width=1.0\columnwidth]{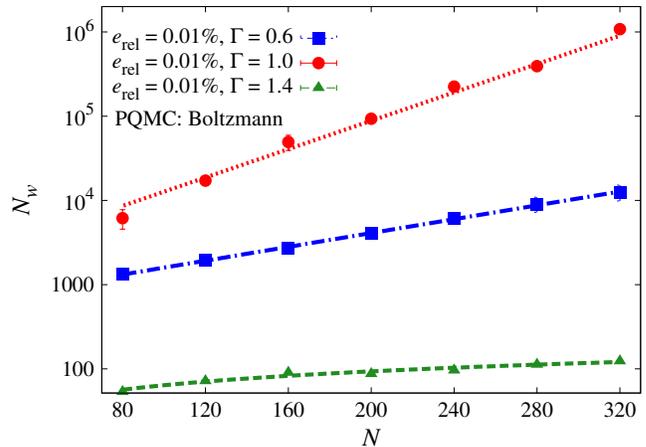}
\caption{(color online). Number of random walkers $N_w$ required to determine, using the optimized Boltzmann-type wave function to guide importance sampling in the 
PQMC simulation, the ground-state energy with a relative error $e_{\mathrm{rel}}$, see Eq.~\eqref{eqn:erel}, as a function of the system size $N$. 
Different datasets correspond to different transverse field intensities $\Gamma$. 
The (red) dotted and (blue) dot-dashed lines represent exponential fitting functions, while the (green) dashed line represents a power-law fit with power $b=0.54(5)$.
}
\label{fig3}
\end{center}
\end{figure}
%%%%%%%%%%%%%%%%%%%%%%%%%%%%%%%%%%%%%%%%%%%%%%%%%%%%%%%%%%%%%%%%%%%%%
%%%%%%%%%%%%%%
%
%%%%%%%%%%%%%%%%%%%%%%%%%%%%%%%%%%%%%%%%%%%%%%%%%%%%%%%%%%%%%%%%%%%%%
%%%   figure 4
%%%%%%%%%%%%%%%%
\begin{figure}
\begin{center}
\includegraphics[width=1.0\columnwidth]{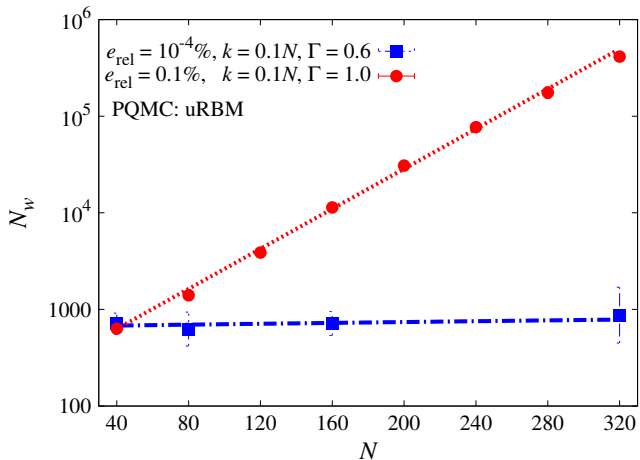}
\caption{(color online).  Number of random walkers $N_w$ required to determine, using the optimized uRBM Ansatz to guide importance sampling in the PQMC simulation, the ground-state energy with a relative error $e_{\mathrm{rel}}$, see Eq.~\eqref{eqn:erel}, as a function of the system size $N$. The number of single-spin Metropolis updates of the hidden spins per CTGFMC hidden-spin update is $k=0.1N$. The (red) dotted line represents and exponential fit, while the (blue) dot-dashed line represents a linear fit.
}
\label{fig4}
\end{center}
\end{figure}
%%%%%%%%%%%%%%%%%%%%%%%%%%%%%%%%%%%%%%%%%%%%%%%%%%%%%%%%%%%%%%%%%%%%%
%%%%%%%%%%%%%%

In the long imaginary-time limit, the walkers sample spin configurations with a probability distribution proportional to 
$f(\X,\tau \rightarrow \infty)=\Psi_{0} ({\X}) \psi_T({\X})$. If $\psi_T({\X})$ is a good approximation of the ground-state wave function, this distribution closely approximates the quantum-mechanical probability of finding the system in the spin configuration $\X$. 
It is important to notice that if our guiding wave function was exact, i.e. if $\psi_T({\X})=\Psi_0({\X})$, then the local energy $E_{\mathrm{loc}}(\X)$ would be a constant function. This would completely suppress the fluctuations of the number of  walkers, therefore eliminating the bias due to the finite walkers population $N_w$. If $\psi_T({\X})$ is, albeit not exact, a good approximation of $\Psi_0({\X})$, the fluctuations of the number of walkers are still reduced compared to the case of the simple CTGFMC algorithm (which corresponds to setting $\psi_T(\X) = 1$) giving a faster convergence 
to the exact $N_w\rightarrow \infty$ limit.
%$\psi_T(\X) = 2^{-N/2}$
%
Below we  consider the use of the variational wave-functions $\Psi_{\bkappa}(\X)$ described in Sec.~\ref{secvar} as guiding wave-functions for the PQMC algorithm, setting the variational parameters $\bkappa$ at their optimal values.

In order to employ the unrestricted neural-network states as guiding functions, the PQMC algorithm has to be modified. One has to implement a combined dynamics of the visible-spin configurations $\X$ and of the hidden-spin configurations $\h$. We will indicate the global configuration as $\xv=(\X,\h)$. The goal is to sample global configurations with the (normalized) probability distribution 
\begin{equation} \label{eqn:py}
p(\xv)=\frac{ \Psi_{0}(\X) \phi_{\bkappa}(\X,\h) }{\sum_{\X,\h}  \Psi_{0}(\X) \phi_{\bkappa}(\X,\h)} \;. 
\end{equation} 
This allows one to compute the ground state energy as 
$E=\lim_{N_c \rightarrow \infty} \sum_{i=1}^{N_c} E_{\loc}( \X_i,\h_i)/N_c$, where $N_c$ is  a number of uncorrelated configurations $\left\{\xv_i\right\}$ sampled from $p(\xv)$.
The local energy $E_{\loc}( \X,\h)$ is defined as in Eq.~\eqref{elocal}. 
A suitable algorithm was implemented in Ref.~\cite{whitlock91} in the case of the continuous-space Green's function Monte Carlo algorithm, where importance sampling was implemented using shadow wave functions. Here we modify the approach of Ref.~\cite{whitlock91} to address quantum spin models.
The visible-spins configurations $\X$ are evolved according to the CTGFMC described above, keeping the hidden-spin configuration $\h'$ fixed. 
The modified imaginary-time Green's function is now 
$\tilde{G}({\X},{\X}^\prime,\Delta \tau|\h')= G(\X,\X^\prime,\Delta \tau)\frac{\phi_{\bkappa}(\X,\h')}{\phi_{\bkappa}(\X',\h')}$. 
As discussed above, this has to be rewritten as the product of a stochastic matrix, which defines how the visible-spin configurations updates are selected, and a weight term, which is taken into account with the branching process.
The weight-update factor is $b_{\xv'}= \sum_{\X} \tilde{G}({\X},{\X}^\prime,\Delta \tau|\h')$.
The dynamics of the hidden-spins configurations is dictated by a (classical) Markov chain Monte Carlo algorithm. 
Considering $\phi_{\bkappa}(\X,\h)$ as an unnormalized probability distribution allows one to write --- for any fixed visible-spin configuration $\X$ --- the Master equation:
\bea
 \phi_{\bkappa}(\X,\h) =\sum_{\h'}  T(\h,\h'|\X)  \phi_{\bkappa}(\X,\h'),
\label{masterEq}
\eea
where $ T(\h,\h'|\X)$ is the transition matrix that defines the Markov process. Clearly, the following condition must be fulfilled $\sum_{\h} T(\h,\h'|\X)=1$, for any $\X$.

Our choice is a single spin flip Metropolis algorithm, where the flip of a randomly selected spin is proposed, and accepted with the probability
\beq
A(\h' \rightarrow \h | \X)= \mathrm{Min} \left\{1,  \frac{ \phi_{\bkappa}(\X,\h)}{ \phi_{\bkappa}(\X,\h')}  \right\}.
\eeq
Here, $\h$ differs from $\h'$ only for the (randomly selected) flipped spin. One could perform a certain number, call it $k$, of Metropolis updates, without modifying the formalism. In fact, this turns out to be useful, as discussed below.
The combined dynamics of the visible and the hidden spins is driven by the following equation:
\begin{equation}
f(\xv,\tau+\Delta \tau) = \sum_{\xv^\prime} G(\xv,\xv^\prime,\Delta \tau) f(\xv^\prime,\tau), 
\end{equation}
with $G(\xv,\xv^\prime,\Delta \tau)= T(\h,\h'|\X)\tilde{G}({\X},{\X}^\prime,\Delta \tau|\h')$.
It can be shown~\cite{whitlock91} that the equilibrium probability distribution of this equation is the desired joint probability distribution 
$p(\xv)$ in Eq.~\eqref{eqn:py}.
The stochastic process corresponding to this equation can be implemented with the following steps: 
\begin{description}
\item[i)] perform the visible-spin configuration update $\X' \rightarrow \X$, keeping $\h'$ fixed, according to the CTGFMC algorithm described above 
(including accumulation of the weight factor); 
\item[ii)] perform $k$ single-spin Metropolis updates of the hidden-spin configuration $\h'$, keeping $\X$ fixed; 
\item[iii)] perform branching of the global configuration.
\end{description}
It is easily shown that the hidden-spin dynamics does not directly affect the weight factor since the normalization of the Green function of the combined dynamics is set by $b_{\xv'}$.

Since the optimized uRBM describes the ground state wave function with high accuracy, one expects that its use as guiding function leads to a drastic reduction of the systematic errors due to the finite random walker population. However, one should take into account that there might be statistical correlations among subsequent hidden-spin configurations along the Markov chain. This might in turn affect the systematic error. Clearly, increasing the number of Metropolis steps $k$ per CTGFMC visible-spin configuration update allows one to suppress such correlations, possibly reducing the systematic error. This will indeed turn out  to be important, in particular at the quantum critical point where statistical correlations along the Markov chain are more significant.

%%%%%%%%%%%%%%%%%%%%%%%%%%%%%%%%%%%%%%%%%%%%%%%%%%%%%%%%%%%%%%%%%%%%%
%%%   figure 5
%%%%%%%%%%%%%%%%
\begin{figure}
\begin{center}
\includegraphics[width=1.0\columnwidth]{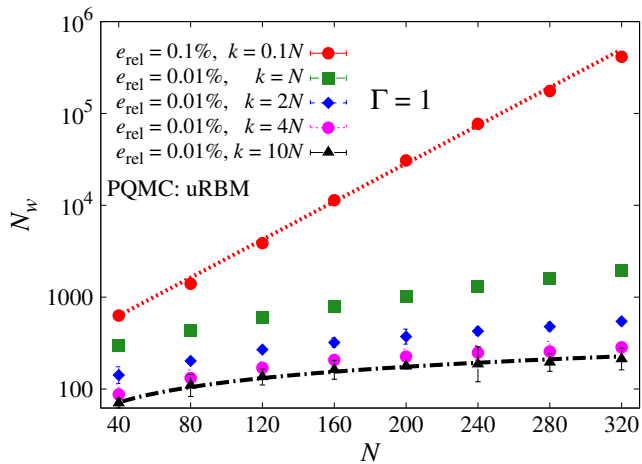}
\caption{(color online). Number of random walkers $N_w$ required to determine, using the optimized uRBM to guide importance sampling in the PQMC simulation, the ground-state energy with a relative error $e_{\mathrm{rel}}$, see Eq.~\eqref{eqn:erel}, as a function of the system size $N$. The transverse field intensity is set at the ferromagnetic quantum critical point  $\Gamma=1$. Different datasets correspond to different values of the the number of single-spin Metropolis updates $k$. The (red) dotted line represents an exponential fit, while the (black) dot-dashed line represents a power-law fit, with power $b=0.55(1)$.
}
\label{fig5}
\end{center}
\end{figure}
%%%%%%%%%%%%%%%%%%%%%%%%%%%%%%%%%%%%%%%%%%%%%%%%%%%%%%%%%%%%%%%%%%%%%
%%%%%%%%%%%%%%

Following Ref.~\onlinecite{inack2}, we analyze the computational complexity of the PQMC algorithm by determining the number of walkers $N_w$ needed to determine  the ground state energy of the Hamiltonian~(\ref{H}) with a prescribed accuracy. All data described below have been obtained with a time step $\Delta \tau = 0.1$, and all simulations have been run for a long enough total imaginary time to ensure equilibration.

First, we consider the simple PQMC algorithm i.e., performed without importance sampling. Fig.~\ref{fig2} displays the scaling with the system size $N$ of the number of walkers $N_w$ required to keep the relative error $e_{\mathrm{rel}}$, defined in Eq.~\eqref{eqn:erel}, at the chosen threshold. 
This scaling is evidently exponential, below, above, and also at the quantum critical point. 
The most severe scaling comes from the ordered phase and could be attributed to the fact that the simple PQMC is formally equivalent to PQMC with a constant $\psi_T(\X)$ for importance sampling. This turns out to be a very poor choice of the guiding function in the ordered regime given that it treats all configurations on an equal footing.
Analogous results have been obtained in Ref.~\onlinecite{inack2} using the diffusion Monte Carlo algorithm. 
This is another PQMC method --- in fact very similar to the CTGFMC algorithm employed here --- whose transition matrix is defined from the imaginary time Green's function derived within the symmetrized Trotter decomposition.
%
% However, the slope at $\Gamma =1$ is expected to be the steepest in the limit when $N_w \rightarrow \infty$~\cite{inack2}.  
%
Introducing importance sampling using the optimized Boltzmann-type Ansatz as guiding function significantly reduces the systematic error due to the finite random walker population, allowing one to reach quite small relative errors. In particular, in the paramagnetic phase at $\Gamma=1.4$, the scaling of $N_w$ versus $N$ is quite flat (see Fig.~\ref{fig3}); it appears to be well described by the power-law $N_w \sim N^b$ with the small power $b=0.54(5)$, rather than by an exponential. However, in the ferromagnetic phase at $\Gamma=0.6$ and at the quantum critical point $\Gamma=1$ the scaling is still clearly exponential. This means that the simple Boltzmann-type Ansatz is, in general, insufficient to ameliorate the exponentially scaling computational cost of the PQMC algorithm.
Fig.~\ref{fig4} shows the scaling of $N_w$ obtained using the optimized uRBM Ansatz as the guiding function. 
The number of hidden-spin Metropolis steps per visible-spin update is set to a (small) fraction of the system size $N$, namely to $k=0.1N$. 
At $\Gamma=0.6$, the required walker population size $N_w$ turns out to be essentially independent on the system size $N$. 
It is worth noticing that the prescribed relative error is here as small as $e_{\mathrm{rel}}=10^{-6}$, and that this high accuracy is achieved with a rather small walkers population $N_w\lesssim 1000$. 
However, at the quantum critical point, $N_w$ still displays an exponential scaling with system size. This effect can be traced back to the diverging statistical correlations among subsequent hidden-spin configurations along the Markov chain, due to quantum criticality. 
As anticipated above, these statistical correlations can be suppressed by increasing the number of hidden-spin updates $k$. 
Fig.~\ref{fig5} displays the scaling of $N_w$, at the quantum critical point, for different $k$ values. 
One observes that the scaling substantially improves already for moderately larger $k$ values, leading to a crossover from the exponential scaling obtained with $k=0.1N$, 
to a square-root like scaling $N_w \sim N^{0.55(1)}$ when $k=10N$. 
It is important to point out that increasing $k$ implies a correspondingly increasing contribution to the global computational cost of the PQMC algorithm. 
However, since $k$ is here linear in the system size, this contribution does not modify, to leading order, the scaling of the global computational cost. 
Therefore, one can conclude that the uRBM Ansatz is sufficient to change the scaling of the computational cost of the PQMC algorithm from exponential in the system size, to an amenable polynomial scaling.
In the simulations presented here, single-spin flip Metropolis updates are employed for the hidden variables. 
It is possible that cluster spin updates would lead to an even faster convergence to the exact $N_w\rightarrow \infty$ limit, due to the more efficient sampling of the hidden-spin configurations. However, such cluster updates cannot always be implemented, in particular for frustrated disordered Hamiltonians relevant for optimization problems; therefore, we do not consider them here.

%%%%%%%%%%%%%%%%%%%%%%%%%%%%%%%%%%%%%%%%%%%%%%%%%%%%%%%%%%%%%%%
\section{Conclusions}
\label{secconc}
The accuracy of variational wave-functions that mimic unrestricted Boltzmann machines, which we refer to as unrestricted neural network states, has been analyzed using the one-dimensional ferromagnetic Ising model as a testbed. By optimizing just three variational parameters, ground-state energies with a relative error smaller than $10^{-3}$ have been obtained. The ferromagnetic quantum phase transition turns out to be the point where the relative error is the largest. This accuracy is comparable to the one previously obtained using restricted neural network states with few hidden variables per visible spin~\cite{carleotroyer}. These restricted neural network states involve a number of variational parameters proportional to the system size, as opposed to the unrestricted neural network states considered here, where the (small) number of variational parameters is fixed. This feature of the unrestricted states makes them very suitable in the context of quantum annealing simulations for Ising-type models (which are sign-problem free). However, since one has to integrate over hidden-spins configurations via Monte Carlo sampling, as opposed to the case of the restricted neural network states~\cite{carleotroyer} --- for which the hidden-spin configurations can be integrated out --- they represent a less promising approach to model ground-states of Hamiltonian where the negative sign-problem occurs. Indeed, in such case an accurate variational Ansatz might have to include also hidden-spins configurations with negative wave-function amplitude, making Monte Carlo integration via random sampling inapplicable.

The variational study summarized here represented a necessary preliminary step to investigate the use of optimized unrestricted neural network states as guiding functions for importance sampling in PQMC simulations. We have found that unrestricted neural network states allow one to drastically reduce  the systematic bias of the PQMC algorithm originating from the finite size of the  random-walker population. Specifically, the scaling of the population size required to keep a fixed relative error as the system size increases changes from the exponential scaling characteristic of simple PQMC simulations performed without guiding functions, to a polynomial scaling. This also implies a corresponding change in the scaling of the computational cost. This qualitative scaling change occurs above, below, and also at the ferromagnetic quantum phase transition. Instead, a conventional variational Ansatz of the Boltzmann type was found to provide a significant improvement of the computational cost only above the critical point (in the paramagnetic phase), but to provide only a marginal improvement at and below the transition.
It is worth emphasizing that the use of unrestricted neural network states as guiding functions in PQMC simulations requires the  sampling of both the visible  and the hidden spins, using the combined algorithm described in Sec.~\ref{secgfmc} (more efficient variants might be possible). The role of the statistical correlations among hidden-spin configurations shows up in particular at the ferromagnetic quantum critical point. We found that these correlations can be eliminated by performing several single-spin updates, still without affecting, to leading order, the global computational complexity of the simulation. 

In Ref.~\cite{bravyi2017polynomial} it was proven that it is possible to devise polynomially-scaling numerical algorithms to determine the ground-state energy, with a small additive error, of various ferromagnetic spin models, including the ferromagnetic Ising chain considered here. However, practical implementations have not been provided. 
The numerical  data we have reported in this manuscript indicate that the PQMC algorithm guided by an optimized unrestricted neural network state represents a practical algorithm  with polynomial computational complexity for the ferromagnetic quantum Ising chain.  
More in general, it was shown in Ref.~\cite{bravyi2015monte} that the problem of estimating
the ground-state energy of a generic sign-problem free Hamiltonian with a
small additive error is at least NP-hard. Indeed, this task encompasses hard optimization problems such as $k-$SAT and MAX-CUT. This suggest that there might be relevant models where the  unrestricted neural network states discussed here are not sufficient to make the computational cost of the PQMC simulations affordable. Relevant candidates are Ising spin-glass models with frustrated couplings. Such systems might require more sophisticated guiding functions obtained, e.g., including more hidden-spin layers in the unrestricted neural network state, as discussed in Sec.~\ref{secvar}.
In future work we plan to search for models that make PQMC simulation problematic. We argue that this will help us in understanding  if and for which models a  systematic quantum speed-up in solving optimization problems using quantum annealing devices, instead of PQMC simulations performed on classical computer, could be achieved.

We acknowledge insightful discussions with  Giuseppe Carleo, Rosario Fazio, Guglielmo Mazzola, Francesco Pederiva, Sandro Sorella, and Matteo Wauters.
S. P. and L. D. acknowledge financial support from the BIRD2016 project ``Superfluid properties of Fermi gases in optical potentials'' of the 
University of Padova. GES acknowledges support by the EU FP7 under ERC-MODPHYSFRICT, Grant Agreement No. 320796. 

%\bibliography{Ref}

\end{document}